\documentclass[british,sigconf,nonacm]{acmart}
\pdfoutput=1

\usepackage{xparse}
\usepackage{xpatch}
\usepackage{etoolbox}
\usepackage{environ}

\usepackage[T1]{fontenc}
\usepackage[utf8]{inputenc}
\usepackage{microtype}

\usepackage{xcolor}

\usepackage[british]{babel}
\usepackage[all]{foreign}
\defasforeign[etsimilia]{et similia}
\defasforeign[Etsimilia]{Et similia}
\defnotforeign[wrt]{{\xperiodafter{w.r.t}}}
\defnotforeign[wl]{{\xperiodafter{w.l.o.g}}}

\usepackage{amsmath}
\usepackage{amsfonts,amssymb,stmaryrd,latexsym}
\usepackage{mathtools}
\usepackage{esvect}

\SetSymbolFont{stmry}{bold}{U}{stmry}{m}{n}

\allowdisplaybreaks

\AtEndPreamble{
	\theoremstyle{acmdefinition}

}

\usepackage{float}
\usepackage{placeins}

\usepackage{booktabs}

\usepackage[inline]{enumitem}

\usepackage{graphicx}
\usepackage{tikz}
\usetikzlibrary{calc,arrows,positioning}

\usepackage{hyperref}
\usepackage{nameref}

\usepackage[capitalise,nameinlink,noabbrev]{cleveref}

\makeatletter
\newcommand{\customlabel}[4][0]{%
	\protected@write\@auxout{}{\string\newlabel{#3}{{#4}{\thepage}{#4}{#3}{}}}%
	\protected@write\@auxout{}{\string\newlabel{#3@cref}{{[#2][#1][#1]#4}{\thepage}}}%
}
\newcommand{\crefv}[1]{%
	\begingroup\@cref@compressfalse\@cref@sortfalse\cref{#1}\endgroup%
}
\newcommand{\Crefv}[1]{%
	\begingroup\@cref@compressfalse\@cref@sortfalse\Cref{#1}\endgroup%
}
\makeatother

\hypersetup{
	hidelinks,
	final,
}

\setcitestyle{numbers,sort&compress}
\usepackage[disable]{todonotes}

\setcopyright{none}

\begin{document}

\title{Towards a Formal Model for Composable Container Systems}

\author{Fabio Burco}
\affiliation{
  \institution{University of Udine}
  \streetaddress{Via delle Scienze 206}
  \city{Udine}
  \postcode{33100}
  \country{Italy} 
}
\email{ }

\author{Marino Miculan}
\orcid{0000-0002-7866-7484}
\affiliation{
  \institution{University of Udine}
  \streetaddress{Via delle Scienze 206}
  \city{Udine}
  \postcode{33100}
  \country{Italy} 
}
\email{marino.miculan@uniud.it}

\author{Marco Peressotti}
\orcid{0000-0002-0243-0480}
\affiliation{
  \institution{University of Southern Denmark}
  \streetaddress{Campusvej 55}
  \city{Odense}
  \postcode{5230}
  \country{Denmark} 
}
\email{peressotti@imada.sdu.dk}

\renewcommand{\shortauthors}{F.~Burco, M.~Miculan, and M.~Peressotti}

\begin{abstract}
In modern cloud-based architectures, \textit{containers} play a central role: they provide powerful isolation mechanisms such that developers can focus on the logic and dependencies of applications while system administrators can focus on deployment and management issue.
In this work, we propose a formal model for container-based systems, using the framework of Bigraphical Reactive Systems (BRSs).
We first introduce \emph{local directed bigraphs}, a graph-based formalism which allows us to deal with localized resources. 
Then, we define a signature for modelling containers and provide some examples of bigraphs modelling containers.
These graphs can be analysed and manipulated using techniques from graph theory: properties about containers can be formalized as properties of the corresponding bigraphic representations.
Moreover, it turns out that the composition of containers as performed by \eg\ \texttt{docker-compose}, corresponds precisely to the composition of the corresponding bigraphs inside an ``environment bigraph'' which in turn is obtained directly from the YAML file used to define the composition of containers.
\end{abstract}



\maketitle

\section{Introduction}

Nowadays, \textit{containers} are increasingly adopted in the design and implementation of cloud-based services. On one hand, containers' isolation mechanisms support a clear separation of tasks, so that developers can focus on the logic and dependencies of applications, while system administrators can focus on deployment and management issues. With a container image, a service can be run on premises as well as on every major cloud provider (such as AWS and Azure). On the other, these isolation mechanisms allow cloud providers to split and dynamically share computing resources according to the a specified multi-tenant model. The flexibility and openness of containerized architectures support easier integration, scalability, dynamic deployment and reconfiguration. 

However, collecting, connecting and coordinating (\emph{orchestrating}) containers into a complete working system is not an easy task. 
Some of the duties of the administrator are: 
provisioning and deployment of containers;
providing each container the resources it needs  (e.g. volumes for file storage);
establishing networks between containers;
exposing services running in a container to the outside world;
remapping port addresses in order to avoid conflicts; etc. 

In order to simplify this task, the administrator is provided with tools and utilities supporting the deployment and management of containers; some examples are Docker's \texttt{docker-compose} and Kubernetes' \texttt{kubectl}.
Still, these tools are complex and error-prone: in their configuration files (for example, \texttt{docker-compose.yml}) it is easy to overlook a required resource, or to mismatch service names, or to misconfigure a network connection hindering communications between containers or (even worse) allowing for unexpected and possibly dangerous communications.  
These problems may be caught at the system's start-up (e.g., a missing image), but may arise suddenly at any time during the system lifetime, causing service malfunctions and interruptions.
In fact, at the moment a system's configuration is defined by trial-and-error: a draft configuration is tested and if errors occur some corrections are made; this operation is repeated several times before the deployment becomes operational, with no guarantee of further misbehaviours.

This situation would benefit from \emph{formal models} of container-based architectures. These models should abstract from the application level details, but still be detailed enough to capture the aspects concerning the composition and connections of containers.
They should allow us to express formally important properties of these systems, and they should support tools for the analysis, verification and manipulation of system's architectures and configurations.

In this work, we propose such a model, within the framework of \emph{Bigraphical Reactive Systems} (BRSs), a family of graph based formalisms introduced as a meta-model for distributed, mobile systems \cite{jm:popl03,milner:bigraphbook,mp:br-tr13}. 
In this approach, system configurations are represented by \emph{bigraphs}, graph-like data structures capable to describe at once both the locations and the logical connections of (possibly nested) components.  
The evolution of a system can be defined in a declarative way by means of a set of \emph{graph rewriting rules}, which can replace and change components' positions and connections.

\begin{figure}
	\centering
	\includegraphics[scale=0.7]{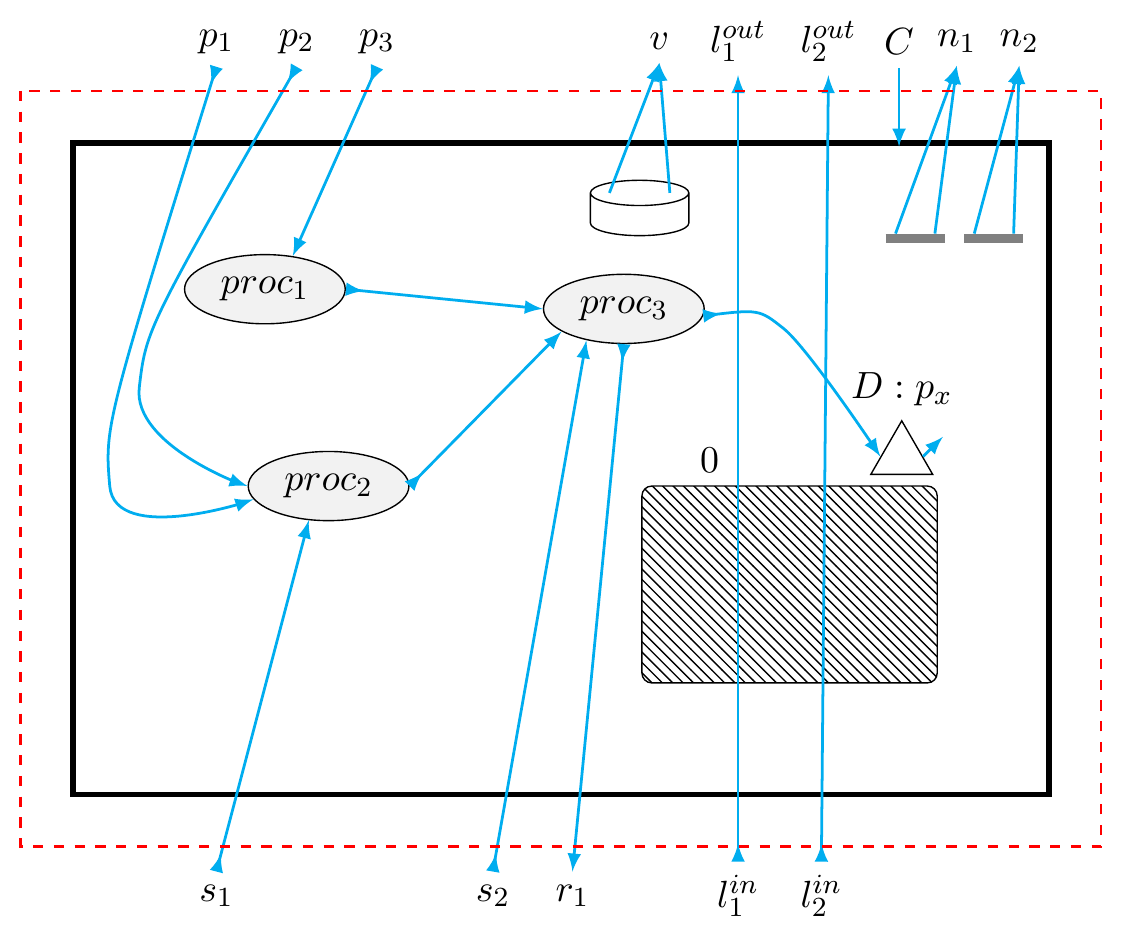}
	\caption{A bigraph representing a container.
		Upper names represent services and resources (volumes, networks) requested or offered to the environment.
	Lower names represent services requested or offered to internal components.}
	\label{fig:example1}
\end{figure}

Like graphs, bigraphs are abstract objects with a precise mathematical formulation, but they have also a simple and accessible graphical representation; see Figure~\ref{fig:example1} for an example.
For this reason, this formalism is a prefect candidate for closing the gap between abstract models for formal methods and tools, and concrete graphical languages for system designers and developers.
Indeed, BRSs have been successfully applied to the formalization of a broad variety of domain-specific models, including context-aware systems and web-service orchestration languages; a non exhaustive list is \cite{jm:popl03,bghhn:coord08,bgm:biobig,dhk:fcm,mmp:dais14,SHB17,sahli:hal-02271394,MSB18}.
Beside their normative power, BRSs are appealing because they provide a range of general results and tools, which can be readily instantiated with the specific model under scrutiny: libraries for bigraph manipulation (e.g., \cite{bgm:dbtk} and \href{http://mads.uniud.it/downloads/libbig/}{jLibBig} \cite{mp:memo14,jlibbig}), simulation tools \cite{mmp:gcm14,mmp:gcm14w}, graphical editors \cite{fph:gcm12}, model checkers \cite{pdh:sac12}, modular composition \cite{pdh:refine11}, \etc. 
Moreover, since bigraphs can be naturally composed, this model allows for \emph{modular design} of container-based services.

The rest of the paper is structured as follows.
In Section~\ref{sec:bigraphs} we introduce \emph{local directed bigraphs}.
In Section~\ref{sec:bigraphsforcontainers} we apply this formalism to model container-based systems, with some examples and showing the correspondence between bigraph composition and container compositions \ala \texttt{docker-compose}.
Some applications of this model to the formalization and verification of security and safety properties about containers are briefly described in Section~\ref{sec:application}.
In Section~\ref{sec:sorting} we briefly mention \emph{sorting disciplines} for ruling out spurious (\ie non well-formed) states, which can be useful for enforcing structural properties of architectures.
Finally, in Section~\ref{sec:concl} we draw some conclusions and sketch directions for further works.

\begin{figure*}[t]
\centering
\includegraphics[scale=0.5]{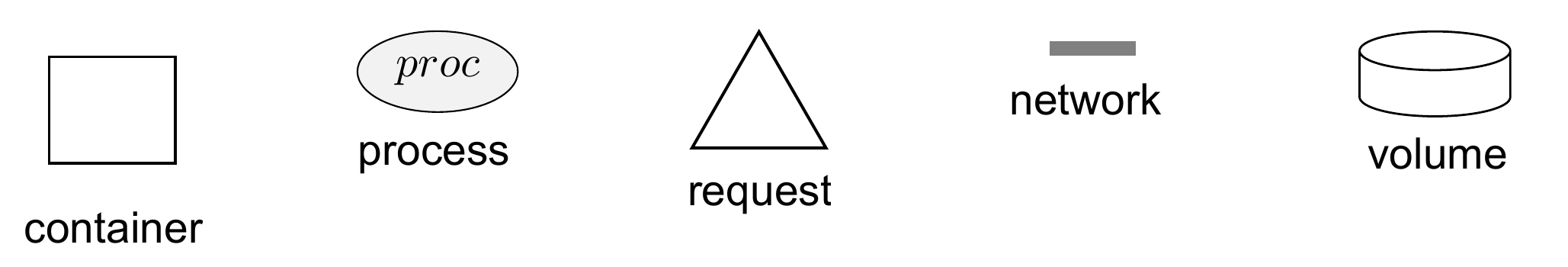}
\caption{Signature for container bigraphs. }
\label{fig:signature}
\end{figure*}

\section{Local directed bigraphs}
\label{sec:bigraphs}
In order to define the formal model for containter-based systems, in this section we introduce \emph{local directed bigraphs}, a variant of directed bigraphs \cite{gm:mfps07} which allows us to deal with \emph{localized} resources.

\begin{definition}
A \emph{(local) interface} is a list $X = \langle X_0, \dots, X_n \rangle $ where each $X_i$ is a pair of disjoint finite sets $(X_i^+, X_i^-)$.
Elements of $X_i^+$, $X_i^-$ are called \emph{positive} (resp. \emph{negative}) \emph{names at $i$}. 
The pair $(X_0^+, X_0^-)$ contains the \emph{global} (i.e., non localized) names.
We define
\[\textstyle
X^+ \triangleq \biguplus_{i=0}^n X_i^+   \qquad
X^- \triangleq \biguplus_{i=0}^n X_i^-    \qquad
width(X) \triangleq n
\]
Two interfaces $X$, $Y$ can be juxtaposed yielding a new interface:
\begin{multline*}
X \otimes Y \triangleq \langle (X_0^+ \uplus Y_0^+, X_0^- \uplus Y_0^-), (X_1^+, X_1^-), \dotsc,\\ (X_n^+, X_n^-), (Y_1^+, Y_1^-), \dotsc, (Y_m^+, Y_m^-) \rangle
\end{multline*}
This operation is associative, and the unit is $\epsilon \triangleq \langle(\emptyset,\emptyset)\rangle$.
\qed
\end{definition}

A \emph{(polarized) signature} $\mathcal{K}$ is a set of elements which form the syntactic basis of bigraphs.
Each type $c \in \mathcal{K}$ is in fact a \emph{polarized arity} $c = \langle n, m \rangle$, for some $n,m\in \mathbb{N}$.

\begin{definition}
Let $\mathcal{K}$ be a signature and $I, O$ two local interfaces.
A \emph{local directed bigraph (ldb)} $B$ from $I$ to $O$, written $B: I \rightarrow O$, is a tuple 
$B = (V, E, ctrl, prnt, link)$ where
\begin{itemize}
	\item $V$ and $E$ are the sets of \emph{nodes} and \emph{edges} respectively;
	\item $ctrl: V \rightarrow \mathcal{K}$ is called the \emph{control map}, and assigns each node an arity;
	\item $prnt: width(I) \uplus V \rightarrow V \uplus width(O)$ is called the \emph{parent map}, and describes the nesting structure of the bigraph (i.e., it has to be a forest); 
	\item $link: Pnt(B) \rightarrow Lnk(B)$ is the \emph{link map}, which describes the directed graph structure, and it is given by the disjoint union of $n+1$ maps
	$link_i: Pnt_i(B) \rightarrow Lnk_i(B)$ such that
	 $\forall x \in link(I_i^+) \cap I_i^- : |link^{-1}(x)|=1$, and $\forall y \in link(O_i^+) \cap O_i^- : |link^{-1}(y)|=1$;
\end{itemize} 
where \emph{ports, points} and \emph{links} are defined as follows, for $i \in \{0, \dotsc, n \}$:
\begin{align*}
Prt_i^+(B) \triangleq {}& \hspace{-1em} \sum_{\substack{v \in V \cr prnt^*(v)=i}} \hspace{-1em} \pi_1(ctrl(v)) \qquad
Prt_i^-(B) \triangleq {} \hspace{-1em} \sum_{\substack{v \in V \cr prnt^*(v)=i}} \hspace{-1em} \pi_2(ctrl(v)) \\
Pnt_i(B) \triangleq {}& I_0^+ \uplus O_0^- \uplus  I_i^+ \uplus O_i^- \uplus Prt_i^+(B) 
\qquad
Pnt  \triangleq {}  \biguplus_{i=1}^n Pnt_i
\\ 
Lnk_i(B) \triangleq {}& I_0^- \uplus O_0^+ \uplus I_i^- \uplus O_i^+ \uplus E \uplus Prt_i^-(B) 
\qquad
Lnk  \triangleq {} \biguplus_{i=1}^n Lnk_i  
\end{align*}

$I$ and $O$ are called \emph{inner} and \emph{outer} interfaces of $B$, respectively.
The localities of the outer interface are called \emph{roots}, while those of the inner interface  \emph{sites}.
\qed
\end{definition}
Notice that, by definition of $prnt$, nodes and sites have to be put under either another node or a root. Hence, differently from names, nodes and sites cannot be global---in fact, there is no root 0.

When interfaces correspond, bigraphs can be ``grafted'' putting roots of one inside sites of the other and connecting links respecting directions as expected.
This yields a definition of \emph{bigraph composition}: for $B_1 : X \rightarrow Y$, $B_2 : Y \rightarrow Z$ two ldb, their composition $B_2 \circ B_1: X \rightarrow Z$ is defined as 
\[
  B_2 \circ B_1 \triangleq (V_1 \uplus V_2, E_1 \uplus E_2, ctrl_1 \uplus ctrl_2, prnt, link)
 \]
where $prnt : width(X) \uplus V \rightarrow V \uplus width(Z)$ and $link : Pnt(B_2 \circ B_1) \rightarrow Lnk(B_2 \circ B_1)$ are defined as expected:
\begin{align*}
prnt(w)=&
\begin{cases}
    prnt_1(w)    & \text{if } w \in width(X) \uplus V \text{ and }\\
                 & \quad  prnt_1(w) \in V_1\\
    prnt_2(prnt_1(w))    & \text{if } w \in width(X) \uplus V \text{ and }\\
                     & \quad prnt_1(w) \in width(Y) \\
    prnt_2(w)    & \text{if } w \in V_2
\end{cases}
\\
link(p)=&
\begin{cases}
prelink(p)    & \text{if } prelink(p) \in Lnk(B_2 \circ B_1) \\
link(prelink(p))    & \text{otherwise}
\end{cases}
\end{align*}
where
\begin{align*}
prelink \triangleq\ & link_1 \uplus link_2  \\
  :\ & Pnt(B_2 \circ B_1) \uplus Y^+ \uplus Y^- \rightarrow Lnk(B_2 \circ B_1) \uplus Y^+ \uplus Y^- \\
Pnt(B_2 \circ B_1) =\ & X^+ \uplus Z^- \uplus Prt^+(B_1) \uplus Prt^+(B_2) \\
Lnk(B_2 \circ B_1) =\ & X^- \uplus Z^+ \uplus Prt^-(B_1) \uplus Prt^-(B_2).
\end{align*}

The identity bigraph on $I$ is $Id_I = \langle\emptyset,\emptyset,\emptyset,id,id\rangle$.

Bigraphs can be combined also by \emph{juxtaposition}, or \emph{tensor product}.
For $B_1 : I_1 \rightarrow O_1$, $B_2 : I_2 \rightarrow O_2$ two ldb,  their product $B_1 \otimes B_2 : I_1 \otimes I_2 \rightarrow O_1 \otimes O_2$ is defined as 
\[
B_1 \otimes B_2 \triangleq (V_1 \uplus V_2, E_1 \uplus E_2, ctrl_1 \uplus ctrl_2, prnt_1 \uplus prnt_2, link_1 \uplus link_2)\text{.}
\]

Therefore, given a signature $\mathcal{K}$, the local interfaces and local directed bigraphs over $\mathcal{K}$ form the objects and arrows of a monoidal category $\mathcal{LDB(K)}$.

\section{Bigraphs for containers}
\label{sec:bigraphsforcontainers}

Bigraphs provide a flexible and immediate framework for designing models and studying their relations in a principled way \cite{milner09:tower}. The main advantage of this approach is that it provides us with a mathematically sound hierarchy of models providing different levels of abstraction to cater to the various analysis and properties of interest. 
We introduce the general approach by defining a model for a possible abstraction level possible model. 
Although it is impossible to fully explore the hierarchy of possible models in the scope of this work, we identify aspects related to compositionality and modularity that is reasonable to expect in any bigraphical model of container systems and show how to capture them using primitives of the framework available in any model.

\FloatBarrier
\subsection{A signature for containers}

Once we have defined the algebraic framework of our model, we can introduce a signature for containers. 
The signature we consider in this paper, and depicted in Figure~\ref{fig:signature}, is the following:
\begin{alignat*}{1}
\mathcal{K} = \{ 
   & container: (0^+, 1^-),
\\ & process_{r, s}: (r^+, s^-),
\\ & request: (1^+, 1^-),
\\ & network: (1^+, 1^-),
\\ & volume: (1^+, 1^-) 
\}
\end{alignat*}
where 
\begin{itemize}
\item
type $container$ is for nodes that represent a container identified by the name connected to its only input port;
\item
type $process_{r, s}$ is for nodes that represent running processes that consumes services connected to its $s$ input ports and offers services over its $r$ output ports (for simplicity we will often omit $r$ and $s$ ans simply write $process$ instead of $process_{r, s}$);
\item
type $request$ is for nodes that represent requests being processes by services implemented by the container;
\item
type $network$ is for nodes that represent network interfaces in the container, network interfaces are connected to form a network through the link graph (see \eg \cref{fig:compose});
\item
type $volume$ is for nodes that represent volumes in the container with linkage providing the associated directory in the host filesystem.
\end{itemize}

These basic elements can be nested and connected, as defined in Section~\ref{sec:bigraphs}, yielding bigraphs such as the one in Figure~\ref{fig:example1}.
Its inner interface is
\[\langle(\emptyset,\emptyset),(\{s_1, s_2,  l_1^{in},l_2^{in}\},\{r_1\})\rangle\] 
and its outer interface is:
\[\langle(\emptyset,\emptyset),(\{v, l_1^{out},l_2^{out},n_1,n_2\},\{p_1,p_2,p_3,C\})\rangle\text{.}\]
This bigraph has one \emph{root}, represented by the red dotted rectangle. Under this root there is one container node, which contains three process nodes, one volume node, two network nodes, a request node, and one \emph{site} (the gray area). 
Arrows connect node ports and names, respecting their polarity. The intended meaning of arrows is that of ``resource accesses'', or dependencies. 
In this example, the container offers services to (i.e., accepts requests from) the surrounding environment on port $p_1,p_2,p_3$, and needs to access a volume $v$ and two networks $n_1,n_2$. 
The site is a ``hole'', meaning that it can be filled by adding another bigraph containing processes (which can access to services offered inside the container through $s_1,s_2$) and resources (which $proc_3$ can access through $r_1$).
Filling a hole with another bigraph corresponds to the composition defined in the previous subsection, and as such it is subject to precise formal conditions, similar to composition of typed functions; in particular, a name of one interface can be connected to that of another interface if their polarity is the same.

\subsection{Composition of containers is composition of bigraphs}
An important example of bigraph composition is represented by the composition of containers, as performed by, e.g., \texttt{docker-compose}. In this case, the context bigraph can be obtained automatically from the \texttt{docker-compose.yml} file.

\begin{figure}[t]
\begin{verbatim}
version: '2'
services:
  wp:
    image: wordpress
    links: [db]
    ports: ["8080:80"]
    networks: [front]
    volumes: [datavolume:/var/www/data:ro]
  db:
    image: mariadb
    expose: ["3306"]
    networks: [front, back]
  pma:
    image: phpmyadmin/phpmyadmin
    links: [db:mysql]
    ports: ["8181:80"]
    volumes: [datavolume:/data]
    networks: [back]
networks:
  front:
    driver: bridge
  back:
    driver: bridge
volumes:
  datavolume:
    external: true
\end{verbatim}
\caption{Example of \texttt{docker-compose.yml} configuration file.}
\label{fig:compose}
\end{figure}

\begin{figure*}
\centering
\begin{tabular}{crlc}
\toprule
(a) & $G_E =$           &
\medskip 
\raisebox{-0.45\height}{\includegraphics[scale=0.85]{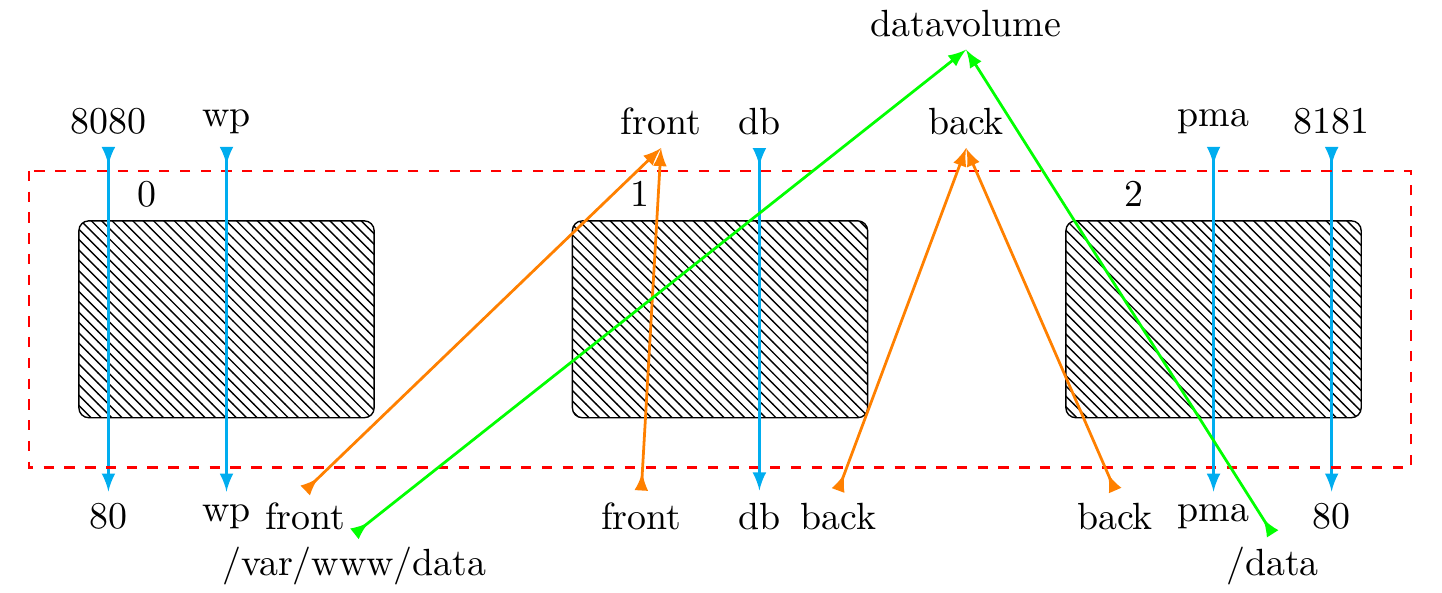}} 
\medskip
& {Environment}
\\
\midrule
(b) & $G_C =$           &
\medskip  \raisebox{-0.45\height}{\includegraphics[scale=0.85]{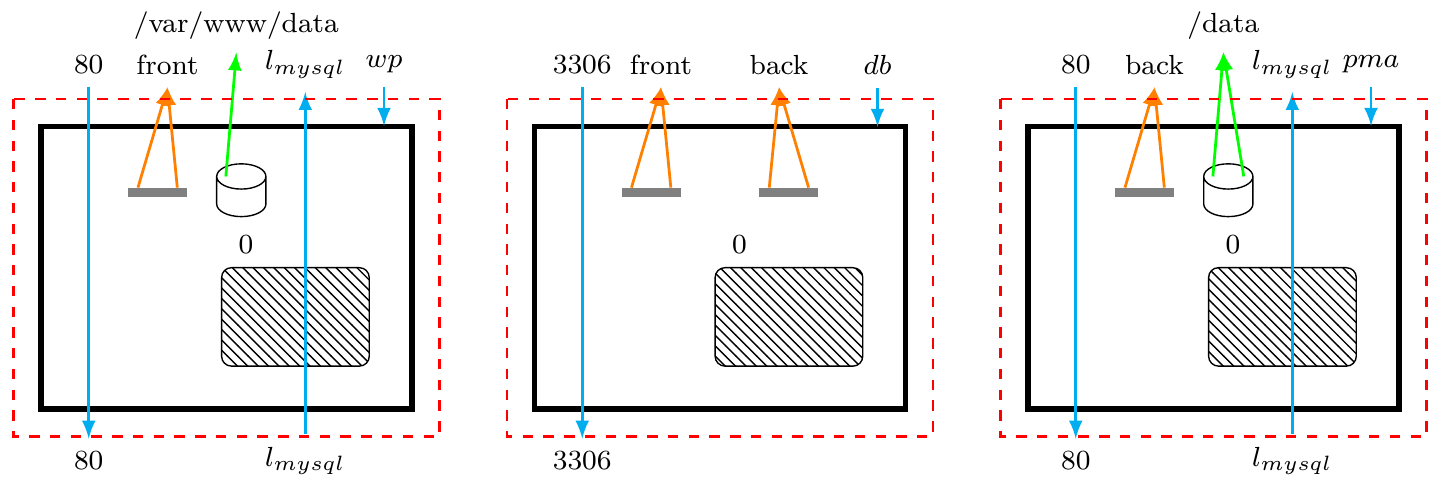}}
\medskip 
& {Containers}
\\
\midrule
(c) & $G_E \circ G_C =$ &
\medskip  \raisebox{-0.45\height}{\includegraphics[scale=0.79]{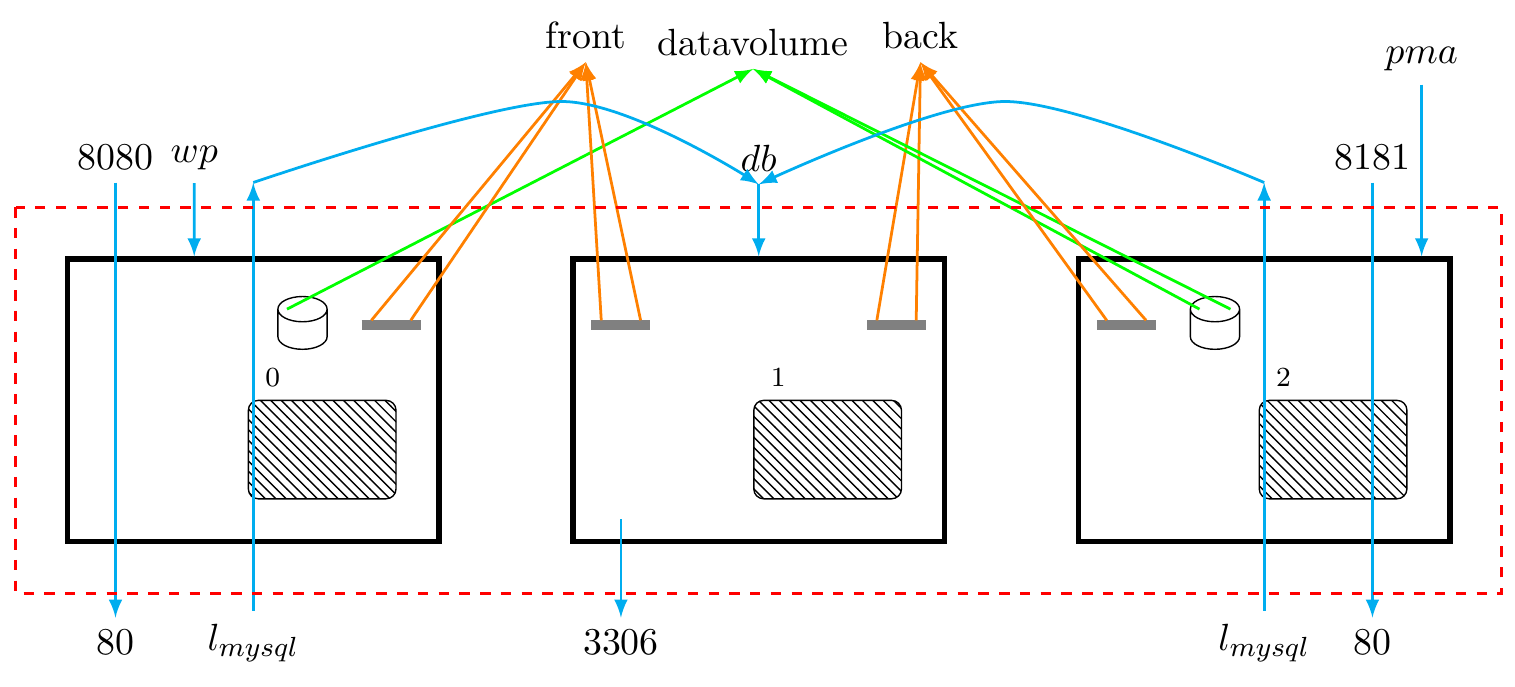}} 
\medskip & {Composite}
\\
\bottomrule
\end{tabular}
\caption{Example of composition:
(a) the composition environment (corresponding to the YAML file in Figure~\ref{fig:compose});
(b) the containers to be assembled;
(c) the result of the composition. }
\label{fig:examplecomp}
\end{figure*}

As an example, let us consider the \texttt{docker-compose.yml} in Figure~\ref{fig:compose}.  Its corresponding ``context'' bigraph is shown in Figure \ref{fig:examplecomp}(a). 
This bigraph has one root (representing the whole resulting system), as many holes as components (``services'') to be assembled, the (possibly shared) networks and volumes that each container requires, and exposes the (possibly renamed) ports to the external environment.  Three bigraphs with the correct interfaces (Figure \ref{fig:examplecomp}(b)) can be composed into the environment, yielding the system in Figure \ref{fig:examplecomp}(c). This resulting system can be seen as a ``pod'', and which can be composed into the site (of the right interface) of other bigraphs, in a modular fashion.

The correspondence between \texttt{docker-compose.yml} files and environment bigraphs can be made formal; in fact, we have implemented a tool which translates \texttt{docker-compose} YAML files into composition bigraphs, taking advantage of the library \href{http://mads.uniud.it/downloads/libbig/}{jLibBig} for bigraph manipulation.

\section{Application: analysis of containers architectures}
\label{sec:application}
Once a container architecture is represented as a bigraph, it can be easily analysed and manipulated.  Many properties about containers can be formalized as properties of the corresponding bigraphic representations and hence verified using well-known techniques from static analysis and model checking.
In this section we exemplify this approach by describing two properties about containers which can be easily verified on their bigraphic representation (and which are not verified by \texttt{docker-compose}).  In fact, we have already implemented a prototype checker to verify  these properties.
 
\subsection{\textit{Links} correctness}
A desirable correctness property about a container configuration is the following: ``no container requires \texttt{links} to containers that cannot be reached through any shared networks''.   Violating this property could lead to run-time misbehaviours, as soon as a container tries to access a service which cannot be reached.  

This property can be easily formalized as a property on the bigraph modeling a container composition (i.e., obtained from a YAML file): for all different containers $C_1,C_2$, if there exists a link from $C_1$ to the (only) port of $C_2$ then there exist a node $N_1$ of type \textsf{network} in $C_1$, a $N_1$ in $C_2$, which are connected to the same name. As an example, the composition in Figure \ref{fig:compose}(c) satisfies this property: $l_{mysql}$ is connected to $db$ and indeed there is a name (``front'') which is connected to by a network in each container $wp$ and $db$; similarly fo $pma$ and $db$ (via ``back'').
This property can be easily verified by a simple reachability check, as implemented in our prototype tool.

\subsection{Network security levels safety}
Let us suppose that networks connecting the containers are ordered according to a \emph{security hierarchy}, specified by the user as a set of ordering assertions of the form $n>m$.  A simple security isolation policy can require that information from (more confidential) network $n$ should not leak to a (lower secure) network $m$---but the flow in the other direction is allowed.
A composition configuration violates this policy if there is a ``channel'', through any number of containers, volumes, and networks, connecting two networks of different security levels against the established order.

We can verify if a given configuration violates this policy by looking for order-violating paths between networks nodes in the corresponding bigraph.
To this end, the bigraph $B$ modelling the system is visited in order to build a bipartite graph $G$ as follows: 
\begin{itemize}
\item nodes of $G$ are each container, network and volume of $B$;
\item for each $c$ container and $r$ a network from which $c$ can read, the arc $(r, c)$ is added to $G$; if $c$ can also write to $r$, then the arc $(c, r)$ is added to $G$. The same applies to volumes.
\end{itemize}
Then, for each pair of nodes $h, l$ such that $h>l$ (in the transitive closure of the order), we check that there is no directed path from $h$ to $l$ in the graph $G$. If this is the case, then the configuration respects the separation policy. Otherwise, an information leakage is possible and hence a warning is raised.

As an example, let us consider again the system in Figure \ref{fig:compose}(c), and let us suppose that ``back $>$ front''. In the corresponding bipartite graph, there is a path from ``front'' to back, via the container $db$.

\section{Sorting}
\label{sec:sorting}
Besides a signature, bigraphical models usually include a \emph{sorting discipline} for ruling out spurious (\ie non well-formed) states. In their more general form sortings are judgments over bigraphs and can be used to model additional properties such as ``only the backend interacts with the database'' in rigorous and declarative ways. 

There are several classes of sortings with specific properties that are of interest for automated verification, in particular \emph{decomposable sortings} \cite{bdh:concur06} and \emph{match sorting} \cite{bgm:dbtk}.
Decomposable sorting are stable under decomposition: if a bigraph is well-sorted then all its decompositions are well-sorted. 
Match sorting disciplines specify sets forbidden bigraphs: a bigraph is well-sorted if it is impossible to find an occurrence (a match) of any of the forbidden bigraphs.

Although decomposable sortings allow for some clever checking algorithms \cite{bdh:concur06,gdbm13:indmatch,gdbh:implmatch}, these are in general reduced in models with low levels of nesting as the ones expected for containers.
Instead, checking if a bigraph is well-sorted according to a match sorting discipline can be done efficiently since the matching problem is fixed-parameter tractable \cite{bmr:tgc14}, and hence for every set of forbidden bigraphs there is an algorithm that can check for occurrences in polynomial time.

\section{Conclusions}
\label{sec:concl}
In this paper we have introduced a bigraphic model of container-based systems.
In this model, containers and container configurations are represented by \emph{bigraphs}, which are graph-like data structures capable to describe at once both the locations and the logical connections of (possibly nested) components of a system. Composition of bigraphs correspond precisely to the composition of containers, as performed, e.g., by \texttt{docker-compose}.
This representation can be used for analysing and verifying properties of container-based. 

Up to our knowledge, this is one of the first formal models for container-based systems. Maybe the closest work is \cite{paraiso2016model}, where a model-driven management of Docker containers has been proposed. This model enables the verification of containers architecture at design time. One difference with our work is that our model is compositional by construction, allowing naturally a modular design and analysis of containers and components.

In this paper we have considered only the static aspects of container based systems.
As future work, we intend to extend our approach to \emph{dynamic} properties of the systems. Evolution of bigraphic systems are usually specified by means of a set of graph-rewriting rules. In our setting, these rewriting rules can be used to model various system reconfigurations, e.g.~scaling up/down or component replacement. Correspondingly, we will be able to take into account other properties, such as safety invariants and liveness properties.

\begin{anonsuppress}
\begin{acks}
	This work was partially supported by the Independent Research Fund Denmark,
	Natural Sciences, grant DFF-7014-00041, 
	and Italian PRIN 2017 grant \emph{IT MATTERS}.
\end{acks}
\end{anonsuppress}

\bibliography{biblio}


\begin{thebibliography}{25}


\ifx \showCODEN    \undefined \def \showCODEN     #1{\unskip}     \fi
\ifx \showDOI      \undefined \def \showDOI       #1{#1}\fi
\ifx \showISBNx    \undefined \def \showISBNx     #1{\unskip}     \fi
\ifx \showISBNxiii \undefined \def \showISBNxiii  #1{\unskip}     \fi
\ifx \showISSN     \undefined \def \showISSN      #1{\unskip}     \fi
\ifx \showLCCN     \undefined \def \showLCCN      #1{\unskip}     \fi
\ifx \shownote     \undefined \def \shownote      #1{#1}          \fi
\ifx \showarticletitle \undefined \def \showarticletitle #1{#1}   \fi
\ifx \showURL      \undefined \def \showURL       {\relax}        \fi
\providecommand\bibfield[2]{#2}
\providecommand\bibinfo[2]{#2}
\providecommand\natexlab[1]{#1}
\providecommand\showeprint[2][]{arXiv:#2}

\bibitem[\protect\citeauthoryear{Bacci, Grohmann, and Miculan}{Bacci
  et~al\mbox{.}}{2009a}]%
        {bgm:biobig}
\bibfield{author}{\bibinfo{person}{Giorgio Bacci}, \bibinfo{person}{Davide
  Grohmann}, {and} \bibinfo{person}{Marino Miculan}.}
  \bibinfo{year}{2009}\natexlab{a}.
\newblock \showarticletitle{Bigraphical models for protein and membrane
  interactions}. In \bibinfo{booktitle}{\emph{Proc.~MeCBIC}}
  \emph{(\bibinfo{series}{Electronic Proceedings in Theoretical Computer
  Science})}, \bibfield{editor}{\bibinfo{person}{Gabriel Ciobanu}} (Ed.),
  Vol.~\bibinfo{volume}{11}. \bibinfo{pages}{3--18}.
\newblock


\bibitem[\protect\citeauthoryear{Bacci, Grohmann, and Miculan}{Bacci
  et~al\mbox{.}}{2009b}]%
        {bgm:dbtk}
\bibfield{author}{\bibinfo{person}{Giorgio Bacci}, \bibinfo{person}{Davide
  Grohmann}, {and} \bibinfo{person}{Marino Miculan}.}
  \bibinfo{year}{2009}\natexlab{b}.
\newblock \showarticletitle{DBtk: a Toolkit for Directed Bigraphs}. In
  \bibinfo{booktitle}{\emph{CALCO 2009 Conference Proceedings - Calco Tools}}
  \emph{(\bibinfo{series}{Lecture Notes in Computer Science})},
  Vol.~\bibinfo{volume}{5728}. \bibinfo{publisher}{Springer}.
\newblock
\newblock
\shownote{Tool available at
  \url{http://sole.dimi.uniud.it/~davide.grohmann/dbtk/}.}


\bibitem[\protect\citeauthoryear{Bacci, Miculan, and Rizzi}{Bacci
  et~al\mbox{.}}{2014}]%
        {bmr:tgc14}
\bibfield{author}{\bibinfo{person}{Giorgio Bacci}, \bibinfo{person}{Marino
  Miculan}, {and} \bibinfo{person}{Romeo Rizzi}.}
  \bibinfo{year}{2014}\natexlab{}.
\newblock \showarticletitle{Finding a Forest in a Tree - The Matching Problem
  for Wide Reactive Systems}. In \bibinfo{booktitle}{\emph{Proc.~TGC}}
  \emph{(\bibinfo{series}{Lecture Notes in Computer Science})},
  \bibfield{editor}{\bibinfo{person}{Matteo Maffei} {and}
  \bibinfo{person}{Emilio Tuosto}} (Eds.), Vol.~\bibinfo{volume}{8902}.
  \bibinfo{publisher}{Springer}, \bibinfo{pages}{17--33}.
\newblock
\showISBNx{978-3-662-45916-4}
\urldef\tempurl%
\url{https://doi.org/10.1007/978-3-662-45917-1_2}
\showDOI{\tempurl}


\bibitem[\protect\citeauthoryear{Birkedal, Debois, and Hildebrandt}{Birkedal
  et~al\mbox{.}}{2006}]%
        {bdh:concur06}
\bibfield{author}{\bibinfo{person}{Lars Birkedal}, \bibinfo{person}{S{\o}ren
  Debois}, {and} \bibinfo{person}{Thomas~T. Hildebrandt}.}
  \bibinfo{year}{2006}\natexlab{}.
\newblock \showarticletitle{Sortings for Reactive Systems}. In
  \bibinfo{booktitle}{\emph{CONCUR}} \emph{(\bibinfo{series}{Lecture Notes in
  Computer Science})}, \bibfield{editor}{\bibinfo{person}{Christel Baier} {and}
  \bibinfo{person}{Holger Hermanns}} (Eds.), Vol.~\bibinfo{volume}{4137}.
  \bibinfo{publisher}{Springer}, \bibinfo{pages}{248--262}.
\newblock
\showISBNx{3-540-37376-4}


\bibitem[\protect\citeauthoryear{Bundgaard, Glenstrup, Hildebrandt,
  H{\o}jsgaard, and Niss}{Bundgaard et~al\mbox{.}}{2008}]%
        {bghhn:coord08}
\bibfield{author}{\bibinfo{person}{Mikkel Bundgaard}, \bibinfo{person}{Arne~J.
  Glenstrup}, \bibinfo{person}{Thomas~T. Hildebrandt}, \bibinfo{person}{Espen
  H{\o}jsgaard}, {and} \bibinfo{person}{Henning Niss}.}
  \bibinfo{year}{2008}\natexlab{}.
\newblock \showarticletitle{Formalizing Higher-Order Mobile Embedded Business
  Processes with Binding Bigraphs}. In
  \bibinfo{booktitle}{\emph{Proc.~COORDINATION}}
  \emph{(\bibinfo{series}{Lecture Notes in Computer Science})},
  \bibfield{editor}{\bibinfo{person}{Doug Lea} {and} \bibinfo{person}{Gianluigi
  Zavattaro}} (Eds.), Vol.~\bibinfo{volume}{5052}.
  \bibinfo{publisher}{Springer}, \bibinfo{pages}{83--99}.
\newblock
\showISBNx{978-3-540-68264-6}


\bibitem[\protect\citeauthoryear{Damgaard, Glenstrup, Birkedal, and
  Milner}{Damgaard et~al\mbox{.}}{2013}]%
        {gdbm13:indmatch}
\bibfield{author}{\bibinfo{person}{Troels~Christoffer Damgaard},
  \bibinfo{person}{Arne~J. Glenstrup}, \bibinfo{person}{Lars Birkedal}, {and}
  \bibinfo{person}{Robin Milner}.} \bibinfo{year}{2013}\natexlab{}.
\newblock \showarticletitle{An inductive characterization of matching in
  binding bigraphs}.
\newblock \bibinfo{journal}{\emph{Formal Aspects of Computing}}
  \bibinfo{volume}{25}, \bibinfo{number}{2} (\bibinfo{year}{2013}),
  \bibinfo{pages}{257--288}.
\newblock


\bibitem[\protect\citeauthoryear{Damgaard, H{\o}jsgaard, and Krivine}{Damgaard
  et~al\mbox{.}}{2012}]%
        {dhk:fcm}
\bibfield{author}{\bibinfo{person}{Troels~Christoffer Damgaard},
  \bibinfo{person}{Espen H{\o}jsgaard}, {and} \bibinfo{person}{Jean Krivine}.}
  \bibinfo{year}{2012}\natexlab{}.
\newblock \showarticletitle{Formal Cellular Machinery}.
\newblock \bibinfo{journal}{\emph{Electronic Notes in Theoretical Computer
  Science}}  \bibinfo{volume}{284} (\bibinfo{year}{2012}),
  \bibinfo{pages}{55--74}.
\newblock


\bibitem[\protect\citeauthoryear{Faithfull, Perrone, and Hildebrandt}{Faithfull
  et~al\mbox{.}}{2013}]%
        {fph:gcm12}
\bibfield{author}{\bibinfo{person}{Alexander~John Faithfull},
  \bibinfo{person}{Gian Perrone}, {and} \bibinfo{person}{Thomas~T.
  Hildebrandt}.} \bibinfo{year}{2013}\natexlab{}.
\newblock \showarticletitle{{BigRed}: A Development Environment for Bigraphs}.
\newblock \bibinfo{journal}{\emph{ECEASST}}  \bibinfo{volume}{61}
  (\bibinfo{year}{2013}).
\newblock


\bibitem[\protect\citeauthoryear{Glenstrup, Damgaard, Birkedal, and
  H{\o}jsgaard}{Glenstrup et~al\mbox{.}}{2007}]%
        {gdbh:implmatch}
\bibfield{author}{\bibinfo{person}{A.J. Glenstrup}, \bibinfo{person}{T.C.
  Damgaard}, \bibinfo{person}{L. Birkedal}, {and} \bibinfo{person}{E.
  H{\o}jsgaard}.} \bibinfo{year}{2007}\natexlab{}.
\newblock \showarticletitle{An Implementation of Bigraph Matching}.
\newblock \bibinfo{journal}{\emph{IT University of Copenhagen}}
  (\bibinfo{year}{2007}).
\newblock
\newblock
\shownote{\url{http://www.itu.dk/~tcd/docs/implBigraphMatching.pdf}.}


\bibitem[\protect\citeauthoryear{Grohmann and Miculan}{Grohmann and
  Miculan}{2007}]%
        {gm:mfps07}
\bibfield{author}{\bibinfo{person}{Davide Grohmann} {and}
  \bibinfo{person}{Marino Miculan}.} \bibinfo{year}{2007}\natexlab{}.
\newblock \showarticletitle{Directed bigraphs}. In
  \bibinfo{booktitle}{\emph{Proc.~MFPS}} \emph{(\bibinfo{series}{Electronic
  Notes in Theoretical Computer Science})}, Vol.~\bibinfo{volume}{173}.
  \bibinfo{publisher}{Elsevier}, \bibinfo{pages}{121--137}.
\newblock


\bibitem[\protect\citeauthoryear{Jensen and Milner}{Jensen and Milner}{2003}]%
        {jm:popl03}
\bibfield{author}{\bibinfo{person}{Ole~H{\o}gh Jensen} {and}
  \bibinfo{person}{Robin Milner}.} \bibinfo{year}{2003}\natexlab{}.
\newblock \showarticletitle{Bigraphs and transitions}. In
  \bibinfo{booktitle}{\emph{POPL}}, \bibfield{editor}{\bibinfo{person}{Alex
  Aiken} {and} \bibinfo{person}{Greg Morrisett}} (Eds.).
  \bibinfo{publisher}{ACM}, \bibinfo{pages}{38--49}.
\newblock
\showISBNx{1-58113-628-5}


\bibitem[\protect\citeauthoryear{Mansutti, Miculan, and Peressotti}{Mansutti
  et~al\mbox{.}}{2014a}]%
        {mmp:gcm14}
\bibfield{author}{\bibinfo{person}{Alessio Mansutti}, \bibinfo{person}{Marino
  Miculan}, {and} \bibinfo{person}{Marco Peressotti}.}
  \bibinfo{year}{2014}\natexlab{a}.
\newblock \showarticletitle{Distributed execution of bigraphical reactive
  systems}.
\newblock \bibinfo{journal}{\emph{{ECEASST}}}  \bibinfo{volume}{71}
  (\bibinfo{year}{2014}).
\newblock
\urldef\tempurl%
\url{https://doi.org/10.14279/tuj.eceasst.71.994}
\showDOI{\tempurl}


\bibitem[\protect\citeauthoryear{Mansutti, Miculan, and Peressotti}{Mansutti
  et~al\mbox{.}}{2014b}]%
        {mmp:dais14}
\bibfield{author}{\bibinfo{person}{Alessio Mansutti}, \bibinfo{person}{Marino
  Miculan}, {and} \bibinfo{person}{Marco Peressotti}.}
  \bibinfo{year}{2014}\natexlab{b}.
\newblock \showarticletitle{Multi-agent Systems Design and Prototyping with
  Bigraphical Reactive Systems}. In \bibinfo{booktitle}{\emph{Distributed
  Applications and Interoperable Systems - 14th {IFIP} {WG} 6.1 International
  Conference, {DAIS} 2014, Held as Part of the 9th International Federated
  Conference on Distributed Computing Techniques, DisCoTec 2014, Berlin,
  Germany, June 3-5, 2014, Proceedings}} \emph{(\bibinfo{series}{Lecture Notes
  in Computer Science})}, \bibfield{editor}{\bibinfo{person}{Kostas Magoutis}
  {and} \bibinfo{person}{Peter~R. Pietzuch}} (Eds.),
  Vol.~\bibinfo{volume}{8460}. \bibinfo{publisher}{Springer},
  \bibinfo{pages}{201--208}.
\newblock
\urldef\tempurl%
\url{https://doi.org/10.1007/978-3-662-43352-2\_16}
\showDOI{\tempurl}


\bibitem[\protect\citeauthoryear{Mansutti, Miculan, and Peressotti}{Mansutti
  et~al\mbox{.}}{2014c}]%
        {mmp:gcm14w}
\bibfield{author}{\bibinfo{person}{Alessio Mansutti}, \bibinfo{person}{Marino
  Miculan}, {and} \bibinfo{person}{Marco Peressotti}.}
  \bibinfo{year}{2014}\natexlab{c}.
\newblock \showarticletitle{Towards distributed bigraphical reactive systems}.
  In \bibinfo{booktitle}{\emph{Proc.~GCM'14}},
  \bibfield{editor}{\bibinfo{person}{Rachid Echahed}, \bibinfo{person}{Annegre
  Habel}, {and} \bibinfo{person}{Mohamed Mosbah}} (Eds.). \bibinfo{pages}{45}.
\newblock
\newblock
\shownote{Workshop version.}


\bibitem[\protect\citeauthoryear{Miculan and Peressotti}{Miculan and
  Peressotti}{2013}]%
        {mp:br-tr13}
\bibfield{author}{\bibinfo{person}{Marino Miculan} {and} \bibinfo{person}{Marco
  Peressotti}.} \bibinfo{year}{2013}\natexlab{}.
\newblock \bibinfo{booktitle}{\emph{Bigraphs Reloaded: a presheaf
  presentation}}.
\newblock \bibinfo{type}{{T}echnical {R}eport} UDMI/01/2013.
  \bibinfo{institution}{Dept.~of Mathematics and Computer Science, Univ.~of
  Udine}.
\newblock
\urldef\tempurl%
\url{http://www.dimi.uniud.it/miculan/Papers/UDMI012013.pdf}
\showURL{%
\tempurl}


\bibitem[\protect\citeauthoryear{Miculan and Peressotti}{Miculan and
  Peressotti}{2014}]%
        {mp:memo14}
\bibfield{author}{\bibinfo{person}{Marino Miculan} {and} \bibinfo{person}{Marco
  Peressotti}.} \bibinfo{year}{2014}\natexlab{}.
\newblock \showarticletitle{A {CSP} implementation of the bigraph embedding
  problem}. In \bibinfo{booktitle}{\emph{Proc.~MeMo}},
  \bibfield{editor}{\bibinfo{person}{Thomas~T. Hildebrandt}} (Ed.),
  Vol.~\bibinfo{volume}{abs/1412.1042}.
\newblock


\bibitem[\protect\citeauthoryear{Miculan and Peressotti}{Miculan and
  Peressotti}{2015}]%
        {jlibbig}
\bibfield{author}{\bibinfo{person}{Marino Miculan} {and} \bibinfo{person}{Marco
  Peressotti}.} \bibinfo{year}{2015}\natexlab{}.
\newblock \bibinfo{title}{jLibBig: a library for bigraphical reactive systems}.
\newblock
\newblock
\urldef\tempurl%
\url{https://github.com/bigraphs/jlibbig}
\showURL{%
\tempurl}


\bibitem[\protect\citeauthoryear{Milner}{Milner}{2009a}]%
        {milner:bigraphbook}
\bibfield{author}{\bibinfo{person}{Robin Milner}.}
  \bibinfo{year}{2009}\natexlab{a}.
\newblock \bibinfo{booktitle}{\emph{The Space and Motion of Communicating
  Agents}}.
\newblock \bibinfo{publisher}{Cambridge University Press}. I--XXI, 1--191
  pages.
\newblock
\showISBNx{978-0-521-73833-0}


\bibitem[\protect\citeauthoryear{Milner}{Milner}{2009b}]%
        {milner09:tower}
\bibfield{author}{\bibinfo{person}{Robin Milner}.}
  \bibinfo{year}{2009}\natexlab{b}.
\newblock \showarticletitle{The tower of informatic models}.
\newblock In \bibinfo{booktitle}{\emph{From semantics to Computer Science;
  Essays in Memory of Gilles Kahn}}. \bibinfo{publisher}{Cambridge University
  Press}.
\newblock


\bibitem[\protect\citeauthoryear{Moudjari, Sahnoun, and Belala}{Moudjari
  et~al\mbox{.}}{2018}]%
        {MSB18}
\bibfield{author}{\bibinfo{person}{Rayene Moudjari},
  \bibinfo{person}{Za{\"{\i}}di Sahnoun}, {and} \bibinfo{person}{Faiza
  Belala}.} \bibinfo{year}{2018}\natexlab{}.
\newblock \showarticletitle{Towards a Fuzzy Bigraphical Multi Agent System for
  Cloud of Clouds Elasticity Management}.
\newblock \bibinfo{journal}{\emph{Int. J. Approx. Reasoning}}
  \bibinfo{volume}{102} (\bibinfo{year}{2018}), \bibinfo{pages}{86--107}.
\newblock
\urldef\tempurl%
\url{https://doi.org/10.1016/j.ijar.2018.07.012}
\showDOI{\tempurl}


\bibitem[\protect\citeauthoryear{Paraiso, Challita, Al-Dhuraibi, and
  Merle}{Paraiso et~al\mbox{.}}{2016}]%
        {paraiso2016model}
\bibfield{author}{\bibinfo{person}{Fawaz Paraiso},
  \bibinfo{person}{St{\'e}phanie Challita}, \bibinfo{person}{Yahya
  Al-Dhuraibi}, {and} \bibinfo{person}{Philippe Merle}.}
  \bibinfo{year}{2016}\natexlab{}.
\newblock \showarticletitle{Model-driven management of docker containers}. In
  \bibinfo{booktitle}{\emph{9th IEEE International Conference on Cloud
  Computing (CLOUD)}}. \bibinfo{pages}{718--725}.
\newblock


\bibitem[\protect\citeauthoryear{Perrone, Debois, and Hildebrandt}{Perrone
  et~al\mbox{.}}{2011}]%
        {pdh:refine11}
\bibfield{author}{\bibinfo{person}{Gian Perrone}, \bibinfo{person}{S{\o}ren
  Debois}, {and} \bibinfo{person}{Thomas~T. Hildebrandt}.}
  \bibinfo{year}{2011}\natexlab{}.
\newblock \showarticletitle{Bigraphical Refinement}. In
  \bibinfo{booktitle}{\emph{Proc.~REFINE}} \emph{(\bibinfo{series}{Electronic
  Proceedings in Theoretical Computer Science})},
  \bibfield{editor}{\bibinfo{person}{John Derrick}, \bibinfo{person}{Eerke~A.
  Boiten}, {and} \bibinfo{person}{Steve Reeves}} (Eds.),
  Vol.~\bibinfo{volume}{55}. \bibinfo{pages}{20--36}.
\newblock


\bibitem[\protect\citeauthoryear{Perrone, Debois, and Hildebrandt}{Perrone
  et~al\mbox{.}}{2012}]%
        {pdh:sac12}
\bibfield{author}{\bibinfo{person}{Gian Perrone}, \bibinfo{person}{S{\o}ren
  Debois}, {and} \bibinfo{person}{Thomas~T. Hildebrandt}.}
  \bibinfo{year}{2012}\natexlab{}.
\newblock \showarticletitle{A model checker for Bigraphs}. In
  \bibinfo{booktitle}{\emph{Proc.~SAC}},
  \bibfield{editor}{\bibinfo{person}{Sascha Ossowski} {and}
  \bibinfo{person}{Paola Lecca}} (Eds.). \bibinfo{publisher}{ACM},
  \bibinfo{pages}{1320--1325}.
\newblock
\showISBNx{978-1-4503-0857-1}


\bibitem[\protect\citeauthoryear{Sahli, Hameurlain, and Belala}{Sahli
  et~al\mbox{.}}{2017}]%
        {SHB17}
\bibfield{author}{\bibinfo{person}{Hamza Sahli}, \bibinfo{person}{Nabil
  Hameurlain}, {and} \bibinfo{person}{Faiza Belala}.}
  \bibinfo{year}{2017}\natexlab{}.
\newblock \showarticletitle{A bigraphical model for specifying cloud-based
  elastic systems and their behaviour}.
\newblock \bibinfo{journal}{\emph{{IJPEDS}}} \bibinfo{volume}{32},
  \bibinfo{number}{6} (\bibinfo{year}{2017}), \bibinfo{pages}{593--616}.
\newblock


\bibitem[\protect\citeauthoryear{Sahli, Ledoux, and Rutten}{Sahli
  et~al\mbox{.}}{2019}]%
        {sahli:hal-02271394}
\bibfield{author}{\bibinfo{person}{Hamza Sahli}, \bibinfo{person}{Thomas
  Ledoux}, {and} \bibinfo{person}{{\'E}ric Rutten}.}
  \bibinfo{year}{2019}\natexlab{}.
\newblock \showarticletitle{{Modeling Self-Adaptive Fog Systems Using
  Bigraphs}}. In \bibinfo{booktitle}{\emph{{FOCLASA 2019 - 17th International
  Workshop on coordination and Self-Adaptativeness of Sotware applications}}}.
  \bibinfo{address}{Oslo, Norway}, \bibinfo{pages}{1--16}.
\newblock
\urldef\tempurl%
\url{https://hal.inria.fr/hal-02271394}
\showURL{%
\tempurl}


\end{thebibliography}

\end{document}